\documentclass{article}

\usepackage[dblblindworkshop,final]{neurips_2026}

\workshoptitle{Autonomous Machine Learning Research}

\usepackage[utf8]{inputenc}
\usepackage[T1]{fontenc}
\usepackage{hyperref}
\usepackage{url}
\usepackage{booktabs}
\usepackage{array}
\usepackage{amsmath}
\usepackage{amssymb}
\usepackage{amsthm}
\usepackage{mathtools}
\usepackage{microtype}
\usepackage{xcolor}
\usepackage{graphicx}
\usepackage{tikz}
\usetikzlibrary{arrows.meta}
\usepackage{enumitem}
\usepackage{float}
\usepackage{flafter}
\usepackage{xspace}
\usepackage[most]{tcolorbox}

\newcolumntype{L}[1]{>{\raggedright\arraybackslash\hyphenpenalty=10000\exhyphenpenalty=10000}p{#1}}

\definecolor{theoremframe}{HTML}{3F6F9F}
\definecolor{theoremfill}{HTML}{F2F7FC}

\newtheorem{theorem}{Theorem}

\theoremstyle{plain}
\newtheorem{definition}{Definition}
\newtheorem{corollary}{Corollary}
\tcbset{resultbox/.style={
  enhanced,
  colback=theoremfill,
  colframe=theoremframe,
  boxrule=0.5pt,
  arc=2pt,
  boxsep=2pt,
  left=4pt,
  right=4pt,
  top=1pt,
  bottom=1pt,
  before skip=6pt,
  after skip=6pt
}}
\tcolorboxenvironment{theorem}{resultbox}
\newtcolorbox{definitionbox}{resultbox}
\newtcolorbox{sourcebox}{enhanced, breakable, colback=black!3, colframe=black!40, boxrule=0.4pt, arc=2pt, left=6pt, right=6pt, top=4pt, bottom=4pt, fontupper=\small, before skip=6pt, after skip=6pt}
\AddToHook{env/sourcebox/begin}{\par\ifdefined\nolinenumbers\nolinenumbers\fi}
\newcommand{\hstar}{H^{\ast}}
\newcommand{\tdeg}{\deg_{\pm}}
\newcommand{\bits}{\{0,1\}}
\newcommand{\signchanges}{C(F)}

\newcommand{\gpt}{\textsc{GPT-5.6 Sol}\xspace}

\title{Formally Verified Autoresearch for Analyzing the Expressivity of Multi-Head Attention}

\author{%
  Karthik Viswanathan$^{1,2}$\thanks{Correspondence to \href{mailto:vkarthik095@gmail.com}{\texttt{vkarthik095@gmail.com}}.} \quad
  Plamen Dimitrov$^{1}$ \quad
  Stefano Rocca$^{1}$ \quad
  Sarayu Sirikonda$^{1,3}$ \\
  $^{1}$Verified Mechanisms \quad $^{2}$MATS Research \quad
  $^{3}$University of California, Berkeley
}

\hypersetup{
  hidelinks,
  pdftitle={Formally Verified Autoresearch for Analyzing the Expressivity of Multi-Head Attention},
  pdfauthor={Karthik Viswanathan, Plamen Dimitrov, Stefano Rocca, Sarayu Sirikonda},
  pdfsubject={NeurIPS 2026 Workshop for Autonomous Machine Learning Research},
  pdfkeywords={mechanistic interpretability, expressivity, formal verification, autonomous research}
}

\begin{document}

\maketitle

\begin{abstract}
Multi-head attention is a central component of the transformer architecture, yet how its expressivity depends on the number of heads remains poorly understood. We study a one-layer, attention-only transformer with additive positional embeddings and define the head complexity $H^\ast(f)$ of a Boolean function $f$ as the minimum number of attention heads required to represent it. We derive an exact linear-fractional normal form for head complexity, prove a general threshold-degree lower bound, and give an explicit ten-bit function for which this bound is strict. For symmetric Boolean functions, we independently show that head complexity equals the number of times the function changes value between consecutive Hamming-weight levels, even when positional embeddings are allowed. This agrees with the concurrent characterization of \citet{hsu2026attentionbased} in a model without positional embeddings. The symmetric lower bound combines our general threshold-degree bound with the classical result of \citet{aspnes1994expressive}. These results were discovered and formalized by a frontier model through conjecture generation, theorem proving, and machine verification, while human researchers asked the high-level questions and reviewed the generated theorems. We make the Lean formalization and research traces publicly available to support agent-assisted theorem proving and formal verification in mechanistic interpretability. This work provides a proof of concept for using the capabilities of frontier models, grounded by formal verification, to develop an understanding of transformer architectures.
\end{abstract}

\noindent\textbf{Keywords:} mechanistic interpretability, expressivity, formal verification, autonomous research.

\section{Introduction}
\label{sec:introduction}

Mechanistic interpretability is an important component of a broader toolkit for AI safety and oversight \cite{ISRSAA2026}. Theoretical mechanistic interpretability works \cite{elhage2021mathematical, elhage2022toymodelssuperposition, geshkovski2025mathematicalperspectivetransformers} have shaped empirical interpretability \cite{olsson2022incontextlearninginductionheads, cunningham2023sparseautoencodershighlyinterpretable, viswanathan2026the} by providing useful abstractions, and principled starting points for experimentation. However, developing this kind of theory has been demanding, requiring sustained expert effort to work through the underlying mathematics. Recent advances in language models for mathematical reasoning and formal theorem proving \cite{tsoukalas2026advancingmathematicsresearchaidriven, alon2026remarksdisproofunitdistance, yang2023leandojo} create an opportunity to accelerate the development of rigorous theory for mechanistic interpretability.

Such theory would be especially valuable for multi-head attention \citep{vaswani2017attention}, since a more fundamental account of how it transforms the residual stream could support mechanistic interpretability in both models at scale and toy models.
In models at scale, a theory of how attention heads transform and combine features could inform theoretically grounded feature-decomposition tools \citep{cunningham2023sparseautoencodershighlyinterpretable, shafran2026from}, by relating the features the probes recover to computations that the architecture can implement. 
For toy models, a theory of multi-head attention can provide a principled basis for interpreting and generalizing empirical findings about how learned computations are distributed across attention heads \citep{nanda2023progressmeasuresgrokkingmechanistic, piotrowski2025constrainedbeliefupdatesexplain}. For instance, the problem studied here arose from an empirical finding in a one-layer transformer trained on modular addition \cite{nanda2023progressmeasuresgrokkingmechanistic}: parity was only weakly linearly decodable from individual head outputs, but became nearly perfectly decodable when the head outputs were combined. This suggested that parity was implemented as a distributed computation across heads, motivating us to ask whether single-head attention faces a structural expressivity barrier.

\paragraph{Related work} Transformer expressivity has been studied through logical characterizations, approximation theory, and analyses of formal languages and Boolean functions \citep{weiss2021thinking,yang2024counting,salzer2026counting,kajitsuka2024one,hahn2020theoretical,kozachinskiy2026parity,hsu2026lowerbounds,kover2026geometry}. \citet{yu2026headcount} establish upper and lower bounds on the parameter complexity required for $\epsilon$-approximation as the number of heads varies, while \citet{hsu2026lowerbounds} establishes lower bounds on the head count for representing $n$-bit parity with one-layer transformers. In concurrent work, \citet{hsu2026attentionbased} also obtain the exact head-count characterization for symmetric functions in a model without positional embeddings. \citet{rajaraman2026headcomplexity} also study head complexity, proving that $n$-bit parity requires exactly $n$ heads and establishing upper and lower bounds for general Boolean functions. Appendix~\ref{app:hsu-comparison} compares these concurrent results with our work.

Despite this progress, deriving general expressivity guarantees for multi-head attention remains technically demanding because tracking even a single layer involves multiple attention head updates and their interactions. Agent-assisted mathematical reasoning can help navigate this analytical complexity, while formal verification provides a rigorous check on the resulting claims. We apply this framework to a concrete expressivity question for Boolean functions:

\begin{quote}
For a Boolean function $f$, what is the minimum number of attention heads needed to represent $f$ in a single attention layer?
\end{quote}

\section{Autonomous Research Result}
\label{sec:autonomous-result}

\label{sec:model}

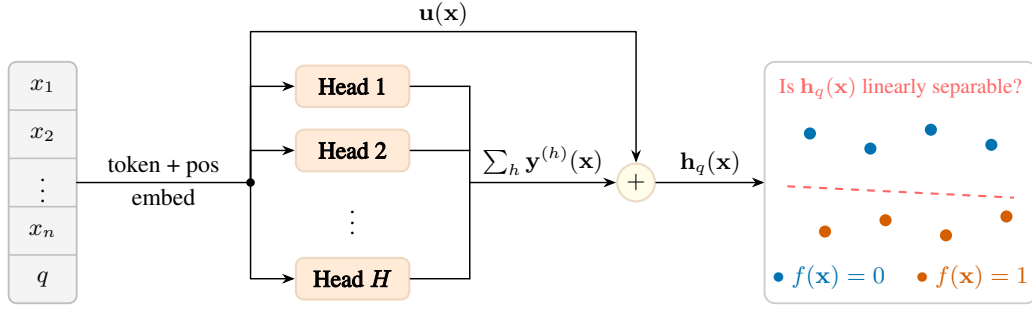
\begin{figure}[tbp]
  \centering
  \noindent
%
%

\providecolor{xorpeach}{HTML}{FFEBD8}
\providecolor{xorpeachline}{HTML}{EFD8C1}
\providecolor{xorcream}{HTML}{FFFDE8}
\providecolor{xorcreamline}{HTML}{ECD9C7}
\providecolor{xorgray}{HTML}{F4F4F4}
\providecolor{xorgrayline}{HTML}{BFBFBF}
\providecolor{xorpanelline}{HTML}{CBCCCB}
\providecolor{xorblue}{HTML}{0072B2}
\providecolor{xororange}{HTML}{D55E00}
\providecolor{xorsalmon}{HTML}{FF6666}

\ifdefined\pdfliteral
  \providecommand{\headlabel}[1]{\pdfliteral direct{q 0 0 0 RG 1 J 1 j 0.2 w 2 Tr}#1\pdfliteral direct{0 Tr Q}}%
\else
  \providecommand{\headlabel}[1]{\textbf{#1}}%
\fi
\begin{tikzpicture}[x=1cm, y=1cm, font=\small,
  block/.style={draw=xorpeachline, fill=xorpeach, rounded corners=3pt,
                line width=0.8pt, align=center, text=black},
  head/.style={block, minimum width=1.5cm, minimum height=0.55cm},
  flow/.style={-{Stealth[length=1.7mm,width=1.3mm]}, black,
               line width=0.55pt, line cap=round, line join=round},
  headflow/.style={black, line width=0.55pt, line cap=round,
                   line join=round},
  note/.style={font=\footnotesize, inner sep=1.5pt},
  classzero/.style={circle, fill=xorblue, minimum size=0.16cm, inner sep=0pt},
  classone/.style={circle, fill=xororange, minimum size=0.16cm, inner sep=0pt}]

  \path[draw=xorgrayline, fill=xorgray, rounded corners=3pt, line width=0.7pt]
    (0,-1.6) rectangle (0.9,1.6);
  \foreach \y in {-0.96,-0.32,0.32,0.96}
    \draw[xorgrayline, line width=0.6pt] (0,\y) -- (0.9,\y);
  \node[text=black] at (0.45,1.28) {$x_1$};
  \node[text=black] at (0.45,0.64) {$x_2$};
  \node[text=black] at (0.45,0) {$\vdots$};
  \node[text=black] at (0.45,-0.64) {$x_n$};
  \node[text=black] at (0.45,-1.28) {$q$};

  \coordinate (split) at (3.2,0);
  \draw[black, line width=0.55pt, line cap=round] (0.9,0) -- (split);
  \node[text=black, anchor=south, inner sep=2pt] at (2.05,0) {token + pos};
  \node[text=black, anchor=north, inner sep=2pt] at (2.05,0) {embed};
  \fill[black] (split) circle (1.6pt);

  \node[head] (headone) at (4.55,1.275) {\headlabel{Head 1}};
  \node[head] (headtwo) at (4.55,0.425) {\headlabel{Head 2}};
  \node[text=black] at (4.55,-0.425) {$\vdots$};
  \node[head] (headh) at (4.55,-1.275) {\headlabel{Head $H$}};

  \draw[flow] (split) |- (headone.west);
  \draw[flow] (split) |- (headtwo.west);
  \draw[flow] (split) |- (headh.west);

  \draw[headflow] (headone.east) -- (6.1,1.275);
  \draw[headflow] (headtwo.east) -- (6.1,0.425);
  \draw[headflow] (headh.east) -- (6.1,-1.275);
  \draw[headflow] (6.1,1.275) -- (6.1,-1.275);

  \node[circle, draw=xorcreamline, fill=xorcream, line width=0.8pt,
        minimum size=0.5cm, inner sep=0pt, font=\small\bfseries]
    (add) at (8.3,0) {$+$};
  \draw[flow] (6.1,0) -- (add.west)
    node[midway, above=1pt, text=black, inner sep=1.5pt] {$\sum_{h} \mathbf{y}^{(h)}(\mathbf{x})$};

  \draw[flow] (split) -- (3.2,2.0) -| (add.north);
  \node[text=black, anchor=south, inner sep=2pt] at (5.75,2.0) {$\mathbf{u}(\mathbf{x})$};

  \draw[flow] (add.east) -- (10.0,0)
    node[midway, above=1pt, text=black, inner sep=1.5pt]
    {$\mathbf{h}_q(\mathbf{x})$};

  \draw[draw=xorpanelline, fill=white, rounded corners=4pt, line width=0.7pt]
    (10.0,-1.6) rectangle (13.6,1.6);
  \node[note, text=xorsalmon] at (11.8,1.25) {Is $\mathbf{h}_q(\mathbf{x})$ linearly separable?};
  \node[classzero] at (10.60,0.65) {};
  \node[classzero] at (11.40,0.45) {};
  \node[classzero] at (12.20,0.7) {};
  \node[classzero] at (13.00,0.5) {};
  \node[classone] at (10.80,-0.65) {};
  \node[classone] at (11.60,-0.5) {};
  \node[classone] at (12.40,-0.7) {};
  \node[classone] at (13.20,-0.45) {};
  \draw[xorsalmon, dashed, line width=0.8pt] (10.3,-0.05) -- (13.3,-0.2);
  \node[text=xorblue, inner sep=1.5pt] at (10.85,-1.25) {$\bullet$\;$f(\mathbf{x})=0$};
  \node[text=xororange, inner sep=1.5pt] at (12.75,-1.25) {$\bullet$\;$f(\mathbf{x})=1$};
\end{tikzpicture}%
  \caption{{\bfseries One-layer attention-only model.} The input consists of $n$ bit tokens followed by a dedicated query token $q$, which also serves as the readout position. Each position is represented by the sum of a token embedding and a positional embedding, and the resulting embedded sequence is shared by all $H$ heads. The heads update all positions in parallel, but the linear readout uses only position $q$. At this position, the head outputs are added to the residual $\mathbf{u}_q$, yielding $\mathbf{h}_q(\mathbf{x})$. Thus, $f$ is represented exactly when the inputs labelled $0$ and $1$ are linearly separable at the readout position $\mathbf{h}_q(\mathbf{x})$, and $\hstar(f)$ is the minimum $H$ for which this occurs.}
  \label{fig:model}
\end{figure}

\paragraph{Setup}
\label{paragraph: setup}
We study a one-layer, attention-only transformer based on multi-head attention \citep{vaswani2017attention}, followed by a linear threshold readout at the position of a dedicated query token $q$, as illustrated in Figure~\ref{fig:model}. For an input $\mathbf{x}$ comprising $n$ bits, let $\mathcal{I}:=[n]\cup\{q\}$ denote the set of sequence positions, and let $\mathbf{u}_j(\mathbf{x})$ denote the input representation at position $j$, obtained by adding its token and positional embeddings. The query representation $\mathbf{u}_q$ is independent of the input because its token and position are fixed. Each head $h$ applies query, key, and value projections $W_Q^{(h)}, W_K^{(h)}, W_V^{(h)}$, followed by an output projection $W_O^{(h)}$ that maps its result back to the model space. Appendix~\ref{app:model-details} gives the full parameterization and dimensions.

The final output at the query position is

\begin{equation}
\mathbf{h}_q(\mathbf{x})=\mathbf{u}_q+\sum_{h=1}^{H}
W_O^{(h)}
\frac{\sum_{j\in\mathcal{I}}
  \exp\!\left(\left\langle W_K^{(h)}\mathbf{u}_j(\mathbf{x}),W_Q^{(h)}\mathbf{u}_q\right\rangle\right)
  W_V^{(h)}\mathbf{u}_j(\mathbf{x})}
{\sum_{j\in\mathcal{I}}
  \exp\!\left(\left\langle W_K^{(h)}\mathbf{u}_j(\mathbf{x}),W_Q^{(h)}\mathbf{u}_q\right\rangle\right)}.
\label{eq:output}
\end{equation}

We say that $f$ is \emph{exactly representable} with $H$ heads if there exist model parameters of the form specified in Appendix~\ref{app:model-details}, and a linear readout $(\mathbf{w},\tau)$ such that

\begin{equation}
\forall\mathbf{x}\in\bits^n,\qquad
\bigl[f(\mathbf{x})=1 \quad\Longleftrightarrow\quad
\left\langle \mathbf{w},\mathbf{h}_q(\mathbf{x})\right\rangle>\tau\bigr].
\label{eq:readout}
\end{equation}

Adding positional embeddings is a seemingly modest but consequential extension: they can break permutation invariance and enlarge the representable function class. This allows us to study nonsymmetric Boolean functions beyond the setting considered by \citet{hsu2026attentionbased}.

\subsection{Characterizing Head Complexity}
The \emph{head complexity} of $f$ is the least number of heads in an exact representation:
\begin{equation}
\hstar(f)=\min\{H: f\text{ is exactly representable by an }H\text{-head model}\}.
\label{eq:hstar}
\end{equation}

Each head contributes one linear-fractional function after readout \citep{charnes1962programming}, giving an equivalent characterization of $\hstar(f)$ without tracking the embeddings and weight matrices in Equations~\eqref{eq:output} and~\eqref{eq:readout}. We call this one-head building block an atom.

\begin{definitionbox}
\begin{definition}[Linear-fractional atom]
\label{def:atom}
A \emph{linear-fractional atom} is a function $\phi:\bits^n\to\mathbb{R}$ that admits a representation
\begin{equation}
\phi(\mathbf{x})=\frac{\mathbf{A}^{\top}\mathbf{x}+b}{\mathbf{C}^{\top}\mathbf{x}+d},
\qquad \mathbf{x}\in\bits^n,
\label{eq:frac-atom}
\end{equation}
where $\mathbf{A},\mathbf{C}\in\mathbb{R}^n$ and $b,d\in\mathbb{R}$ satisfy all three conditions:
\begin{enumerate}[label=(\roman*),leftmargin=*,nosep]
  \item $\mathbf{C}^{\top}\mathbf{x}+d>0$ for every $\mathbf{x}\in\bits^n$.
  \item The entries of $\mathbf{C}$ are all strictly positive, all strictly negative, or all zero.
  \item If $\mathbf{C}=\mathbf{0}$, the entries of $\mathbf{A}$ are likewise all strictly positive, all strictly negative, or all zero.
\end{enumerate}
We call such a coefficient tuple $(\mathbf{A},b,\mathbf{C},d)$ \emph{admissible}. The intercept $b$ is unrestricted, and when $\mathbf{C}\ne\mathbf{0}$ the numerator slopes $\mathbf{A}$ are unrestricted as well.
\end{definition}
\end{definitionbox}

\smallskip

\begin{definition}[Linear-fractional complexity]
\label{def:lfrac}The \emph{linear-fractional complexity} of $f$, denoted $L_{\mathrm{frac}}(f)$, is the least nonnegative integer $H$ for which there exist linear-fractional atoms $\phi_1,\ldots,\phi_H$ and a constant $c\in\mathbb{R}$ such that, for every $\mathbf{x}\in\bits^n$, $c+\sum_{h=1}^{H}\phi_h(\mathbf{x})>0 \Longleftrightarrow f(\mathbf{x})=1$.
\end{definition}
\smallskip

\begin{theorem}[Linear-fractional normal form]
\label{thm:normal-form}
Fix $H\in\mathbb{Z}_{\geq0}$. A function $S:\bits^n\to\mathbb{R}$ is the scalar score $S(\mathbf{x})=\langle\mathbf{w},\mathbf{h}_q(\mathbf{x})\rangle-\tau$ of an $H$-head model in our setting, with unrestricted model and head dimensions, if and only if it has the form
\begin{equation}
S(\mathbf{x})=c+\sum_{h=1}^{H}
\frac{\mathbf{A}_h^{\top}\mathbf{x}+b_h}
{\mathbf{C}_h^{\top}\mathbf{x}+d_h},
\qquad \mathbf{x}\in\bits^n,
\label{eq:score-normal-form}
\end{equation}
where $c\in\mathbb{R}$ and each tuple $(\mathbf{A}_h,b_h,\mathbf{C}_h,d_h)$ is admissible in the sense of Definition~\ref{def:atom}. Each head contributes one atom, and any collection of $H$ atoms can be realized jointly by $H$ heads using shared token and positional embeddings. Consequently, for every Boolean function $f$,
\begin{equation*}
\hstar(f)=L_{\mathrm{frac}}(f).
\end{equation*}
\end{theorem}

An atom thus gives a compact, exact representation of a head's contribution after scalar readout. The full model's score is the sum of these atoms and the constant $c$, which accounts for the fixed query residual and readout threshold.

\subsubsection{Why the correspondence holds}
For notational simplicity, let us fix a head and suppress its head index. Let $\mathbf{e}_0$ and $\mathbf{e}_1$ be the token embeddings for bits zero and one, and let $\mathbf{p}_i$ be the positional embedding at input position $i$. With $\Delta\mathbf{e}=\mathbf{e}_1-\mathbf{e}_0$, the input vector at position $i$, before attention, is
\begin{equation*}
\mathbf{u}_i(\mathbf{x})=\mathbf{e}_{x_i}+\mathbf{p}_i
=\mathbf{e}_0+\mathbf{p}_i+x_i\Delta\mathbf{e}.
\end{equation*}
Thus $\mathbf{u}_i(\mathbf{x})$ combines the token and positional embeddings and depends on the input $\mathbf{x}$ only through its $i$-th bit $x_i$.

Using this head's key and query projections, $W_K$ and $W_Q$, the logit for attending from the query token to position $i$ is
\begin{equation*}
\ell_i(\mathbf{x})
=\left\langle W_K\mathbf{u}_i(\mathbf{x}),W_Q\mathbf{u}_q\right\rangle
=\left\langle\mathbf{u}_i(\mathbf{x}),W_K^{\top}W_Q\mathbf{u}_q\right\rangle.
\end{equation*}
The vector $W_K^{\top}W_Q\mathbf{u}_q$ is fixed as the input varies: the model weights are fixed, and the query representation $\mathbf{u}_q$ is input-independent. The logit is therefore a dot product of $\mathbf{u}_i(\mathbf{x})$ with a fixed vector, making it linear in $\mathbf{u}_i(\mathbf{x})$. Likewise, applying the value projection $W_V$, output projection $W_O$, and scalar readout $\mathbf{w}$ gives $s_i(\mathbf{x})=\langle\mathbf{w},W_O W_V\mathbf{u}_i(\mathbf{x})\rangle$. This is also linear in $\mathbf{u}_i(\mathbf{x})$ because all three maps are linear. Substituting the expression for $\mathbf{u}_i(\mathbf{x})$ gives
\begin{equation*}
\ell_i(\mathbf{x})=\kappa_i+a x_i,
\qquad
s_i(\mathbf{x})=m_i+\beta x_i.
\end{equation*}
Here $\kappa_i$ and $m_i$ are the logit and scalar value when $x_i=0$, while $a$ and $\beta$ are their changes when the bit switches to one. These changes are shared across positions because switching any input bit adds the same vector $\Delta\mathbf{e}$ to its embedding.

\paragraph{Weighted scalar values (numerator).}
Exponentiating the logit gives the unnormalized attention weight $\exp(\ell_i(\mathbf{x}))=\rho_i\alpha^{x_i}$, where $\rho_i=\exp(\kappa_i)>0$ and $\alpha=\exp(a)>0$. Multiplying this weight by the corresponding scalar value $s_i(\mathbf{x})=m_i+\beta x_i$ gives $\rho_i\alpha^{x_i}(m_i+\beta x_i)$. This product is affine in $x_i$, using $\alpha^{x_i}=1+(\alpha-1)x_i$ and $x_i^2=x_i$ on Boolean inputs. Let $\ell_q$ and $s_q$ denote the query token's logit and scalar value, both of which are constant because $\mathbf{u}_q$ is fixed. Summing the weighted scalar values over the input positions and the query token gives the affine numerator
\begin{equation*}
N(\mathbf{x})=\exp(\ell_q)s_q+\sum_{i=1}^{n}\rho_i\alpha^{x_i}(m_i+\beta x_i)
=\mathbf{A}^{\top}\mathbf{x}+b.
\end{equation*}

\paragraph{Unnormalized attention weights (denominator).}
The denominator sums the same unnormalized attention weights over the input positions and the query token. Since $\alpha^{x_i}=1+(\alpha-1)x_i$ on Boolean inputs, this sum is also affine:
\begin{equation*}
D(\mathbf{x})=\exp(\ell_q)+\sum_{i=1}^{n}\rho_i\alpha^{x_i}
=\mathbf{C}^{\top}\mathbf{x}+d.
\end{equation*}
This denominator is strictly positive because every unnormalized attention weight is positive.

The head's scalar contribution is obtained by dividing the weighted scalar-value sum by the total attention weight. Combining the two expressions therefore gives the linear-fractional form of Definition~\ref{def:atom}:
\begin{equation*}
\phi(\mathbf{x})=\frac{N(\mathbf{x})}{D(\mathbf{x})}
=\frac{\mathbf{A}^{\top}\mathbf{x}+b}{\mathbf{C}^{\top}\mathbf{x}+d}.
\end{equation*}

Conversely, every admissible tuple determines valid attention weights and values. With unrestricted model width, separate coordinate blocks allow all $H$ heads to realize their prescribed atoms within the same shared embeddings. Appendix~\ref{app:proof-normal-form} gives both constructions and proves the exact coefficient characterization.

\begin{corollary}[One head]
\label{cor:one-head}
$\hstar(f)=0$ if and only if $f$ is constant, and $\hstar(f)\leq1$ if and only if $f$ is a linear threshold function: there exist reals $\theta_0,\ldots,\theta_n$ such that, for every $\mathbf{x}\in\bits^n$, $f(\mathbf{x})=1 \Longleftrightarrow \theta_0+\sum_i\theta_ix_i>0$.
\end{corollary}
Appendix~\ref{app:proof-one-head} gives the proof. Parity is not a linear threshold function, so Corollary~\ref{cor:one-head} already gives a structural reason why parity was not linearly decodable from single heads in the modular-addition model of Section~\ref{sec:introduction}.

We next relate head complexity to threshold degree. Let $\tdeg(f)$ be the minimum degree of a real polynomial that sign-represents\footnote{A real polynomial $P$ \emph{sign-represents} $f$ if, for every $\mathbf{x}\in\bits^n$, $P(\mathbf{x})>0 \Longleftrightarrow f(\mathbf{x})=1$.} $f$ on the Boolean cube \citep{minsky1969perceptrons}.

\begin{theorem}[Threshold-degree lower bound]
\label{thm:threshold}
Every Boolean function satisfies
\begin{equation*}
\tdeg(f)\leq\hstar(f).
\end{equation*}
\end{theorem}
By Theorem~\ref{thm:normal-form}, an $H$-head score is a constant plus a sum of $H$ linear-fractional atoms. Multiplying by the product of their positive affine denominators preserves its sign pattern on the Boolean cube and yields a polynomial of degree at most $H$. Appendix~\ref{app:proof-threshold} gives the full argument.

\paragraph{Strictness of the lower bound.}
The threshold-degree bound is not always tight. For a ten-bit input $\mathbf{x}=(\mathbf{a},\mathbf{b})\in\bits^{10}$, where $\mathbf{a},\mathbf{b}\in\bits^5$, define
\begin{equation}
\begin{gathered}
Q(\mathbf{a},\mathbf{b})=
\left(\sum_{i=1}^{5}(1-2a_i)\right)\left(\sum_{j=1}^{5}(1-2b_j)\right)
-3\sum_{i=1}^{5}\left(1-2a_i\right)\left(1-2b_i\right), \\
f_{10}(\mathbf{x})=1 \Longleftrightarrow Q(\mathbf{a},\mathbf{b})>0
\end{gathered}
\label{eq:f10}
\end{equation}
\begin{theorem}[Strict separation from threshold degree]
\label{thm:separation}
The function $f_{10}$ satisfies
\begin{equation*}
\tdeg(f_{10})=2<3\leq\hstar(f_{10}).
\end{equation*}
\end{theorem}
Appendix~\ref{app:proof-separation} gives the proof. We found autoresearch particularly useful for constructing examples when the bound in Theorem~\ref{thm:threshold} is strict, resulting in the example above.

\paragraph{Constructive upper bounds.}
We next derive a constructive upper bound for functions that depend on a scalar statistic $t_\lambda(\mathbf{x})$, in terms of the number of distinct values that $t_\lambda$ takes on the Boolean cube.

\begin{theorem}[Weighted-sum interpolation]
\label{thm:weighted-upper}
Let $\lambda_1,\ldots,\lambda_n>0$ and define $t_\lambda:\bits^n\to\mathbb{R}$ by $t_\lambda(\mathbf{x})=\sum_{i=1}^{n}\lambda_i x_i$. Suppose $f$ depends on its input only through $t_\lambda$, so that $f(\mathbf{x})=F(t_\lambda(\mathbf{x}))$ for some function $F:\mathbb{R}\to\bits$. Then
\begin{equation*}
\hstar(f)\leq |\operatorname{Im}(t_\lambda)|-1.
\end{equation*}
\textnormal{For $\lambda_i=2^{i-1}$, we have $|\operatorname{Im}(t_\lambda)|=2^n$, so $\hstar(f)\leq2^n-1$ for every Boolean function $f$.}
\end{theorem}
The proof uses $|\operatorname{Im}(t_\lambda)|-1$ heads to realize the labels assigned by $F$ to the attainable values of $t_\lambda$. Appendix~\ref{app:proof-weighted} gives the construction.

\subsection{Exact Head Complexity for Symmetric Functions}

Let $f:\bits^n\to\bits$ be symmetric. Equivalently, there is a unique function $F:\{0,\ldots,n\}\to\bits$ such that, for every $\mathbf{x}\in\bits^n$, $f(\mathbf{x})=F(|\mathbf{x}|)$, where $|\mathbf{x}|=\sum_{i=1}^{n}x_i$ \citep{paturi1992degree}. We call $F$ the \emph{Hamming-weight profile} of $f$ and define the number of changes in this profile by
\begin{equation}
\signchanges=\left|\{k\in[n]:F(k-1)\neq F(k)\}\right|.
\label{eq:sign-changes}
\end{equation}
Taking $\lambda_i=1$ in Theorem~\ref{thm:weighted-upper} gives $\hstar(f)\leq n$. The following result sharpens this bound to the number of changes in $F$.

\begin{theorem}[Exact symmetric characterization]
\label{thm:symmetric}
For every symmetric Boolean function $f$ with Hamming-weight profile $F$,
\begin{equation*}
\hstar(f)=\tdeg(f)=\signchanges.
\end{equation*}
\end{theorem}

\begin{corollary}[Parity]
\label{cor:parity}
For every $n\geq1$, $\hstar(\mathrm{PARITY}_n)=n$. This follows from Theorem~\ref{thm:symmetric}, since parity has $n$ sign changes across Hamming-weight levels.
\end{corollary}

\paragraph{Lower bound.}
Lemma~2.6 of \citet{aspnes1994expressive} shows that the threshold degree of a symmetric Boolean function equals the number of changes in its univariate Hamming-weight representation, so $\tdeg(f)=\signchanges$. Combining this with Theorem~\ref{thm:threshold} gives $\hstar(f)\geq\signchanges$.

\paragraph{Upper bound.}
We construct a degree-$\signchanges$ polynomial with sign pattern $F$ by placing one root between each pair of consecutive Hamming-weight levels at which $F$ changes. After dividing this polynomial by a positive polynomial of the same degree, a partial-fraction decomposition yields a representation using $\signchanges$ linear-fractional atoms, so Theorem~\ref{thm:normal-form} gives $\hstar(f)\leq\signchanges$. Appendix~\ref{app:proof-symmetric} gives the full construction.

\subsection{Machine-Checked Formalization}

All results above are formalized and machine-checked in Lean~4. The codebase lays the groundwork for a publicly available Lean environment supporting agent-assisted research in mechanistic interpretability. Section~\ref{sec:system} describes the agents' contributions and the human review process.

\section{System Design and Meta-Analysis}
\label{sec:system}

The theorems in Section~\ref{sec:autonomous-result} were derived by \gpt{} from a /goal prompt. The separation\footnote{A \emph{separation} is an explicit Boolean function where the threshold degree is strictly smaller than its head complexity, which shows that the lower bound of Theorem~\ref{thm:threshold} is not always tight.} in Theorem~\ref{thm:separation} was obtained through a separate counterexample search initiated by the authors. The Lean formalizations were generated through straightforward prompting, and human researchers subsequently refined the Lean code to make it legible and reusable. This section is based on the anonymized record of the autonomous run, released as \texttt{theorems-rollout.jsonl} in the repository accompanying this paper.\footnote{The results had already been obtained in an earlier run. For clarity, this section describes the rollout of a subsequent run conducted without access to the earlier run, although its main observations apply to both.}

\subsection{Agent and harness}

The theorems in Section~\ref{sec:autonomous-result} were generated by a single \gpt agent running in a Codex Desktop sandbox. Section~\ref{subsec:tools} describes the execution environment. The agent received the problem setup from Section~\ref{paragraph: setup} and a comprehensive set of acceptance criteria that provided general research guidance without revealing the expected results. In approximately 9.5 hours, it derived the linear-fractional normal form (Theorem~\ref{thm:normal-form}), the threshold-degree lower bound (Theorem~\ref{thm:threshold}), and the exact symmetric characterization (Theorem~\ref{thm:symmetric}). The theorem statements, derivations, and supporting constructions were recorded in \texttt{answer.md}, the run's primary research output. The agent also explored broader bounds not reported in Section~\ref{sec:autonomous-result}. A separate counterexample search subsequently found a 128-bit separation between threshold degree and head complexity, which was later improved to the explicit ten-bit example in Theorem~\ref{thm:separation}.

\paragraph{Defining the problem.}
Considerable human effort went into specifying the research question precisely and fixing a model against which proposed results could be audited. Appendices~\ref{app:problem-statement} and~\ref{app:model-spec} reproduce the problem statement and model specification, respectively. Since a closed-form characterization of $\hstar(f)$ for arbitrary Boolean functions could not be assumed, the problem statement identified several broader forms of progress, including lower and upper bounds, comparisons with classical complexity measures, and exact results for natural function families.

\paragraph{The theorem-generation prompt.}
The prompt, adapted from \citet{openai2026cdcprompt}, asked the agent to conduct a diversified search across multiple proof strategies. It was submitted through Codex's \texttt{/goal} command, which maintains a persistent objective and completion condition across turns \citep{pathak2026goals}. The prompt instructed the agent to explore genuinely different proof strategies, organize them by their underlying mathematical ideas, and support them with concrete equations or counterexamples. Public web search was disabled so that the agent could not retrieve existing solutions during the run. The run was one-shot: after receiving the initial prompt, the agent completed the search and returned its answer without further human input. Appendix~\ref{app:prompt} reproduces the prompt verbatim.

The run used a single \texttt{gpt-5.6-sol} process at \texttt{xhigh} reasoning effort in Codex Desktop. Session metadata records Codex CLI build \texttt{0.148.0-alpha.21}, source \texttt{vscode}, a 258,400-token context window, and four context compactions. Although the multi-agent runtime was available, the agent did not spawn any subagents. It used no external connectors and did not invoke Lean during the initial research interval. Lean was introduced only after a subsequent user request because the initial run focused on developing the theory in natural language.

\paragraph{The \texttt{/goal} command alone was insufficient.}
In preliminary runs, supplying only a \texttt{/goal} objective led the agent to stop after a short period, make only incremental progress, and mark the goal as complete prematurely. Adapting the persistence and search instructions from \citet{openai2026cdcprompt} was therefore important for sustaining a longer research run.

\paragraph{Counterexample-search setup.}
The counterexample search used a separate Codex Desktop session with a single \gpt agent running at \texttt{ultra} reasoning effort. Its first turn comprised 32 tool calls completed in under seven minutes. Section~\ref{subsec:human-interventions} describes how the search was initiated and refined.

\subsection{Tools}
\label{subsec:tools}

The run used Python~3, NumPy, SciPy, SymPy, and standard shell and Git tools for numerical experiments, symbolic and exact calculations, repository inspection, and testing. Table~\ref{tab:tool-usage} summarizes the recorded shell activity. Based on command syntax, we classified 373 of the 387 calls and left the remaining 14 unclassified rather than infer their purpose from context. The empirical checks tested the main constructions in \texttt{answer.md} using exact arithmetic and exhaustive evaluation on small instances; the other classified calls included additional checks on small Boolean functions and the proposed upper-bound constructions.

\begin{table}[H]
  \caption{Recorded shell activity during the theorem-generation run.}
  \label{tab:tool-usage}
  \centering
  \small
  \begin{tabular}{lr}
    \toprule
    Command category & Calls \\
    \midrule
    Python invocations & 153 \\
    \texttt{sed} inspections & 70 \\
    \texttt{rg} searches & 40 \\
    Runs of empirical checks & 24 \\
    Numbered file inspections & 17 \\
    Git queries & 14 \\
    Other classified calls & 55 \\
    Unclassified calls & 14 \\
    \midrule
    Total & 387 \\
    \bottomrule
  \end{tabular}
\end{table}

\paragraph{Sandboxing details.} The run executed in a Codex Desktop sandbox on a local Linux virtual machine: a managed permission profile with read access to the whole file system, write access only to the project directory and temporary directories, and no network access. Any action outside these bounds would have required the user's approval, and the rollout records no such request before delivery. Its computations were exact or small-scale numerical checks.

\paragraph{Details on the counterexample search.} The counterexample search that reduced the initial 128-bit construction to the ten-bit example in Theorem~\ref{thm:separation} used Python and exact solvers for small Boolean instances, with public web search available. The formalization was carried out in Lean~4 using Mathlib and Lake, with computationally intensive builds run on an institutional SLURM cluster.

\subsection{Research loop}

The run proceeded in two stages. The agent spent approximately 9 hours and 38 minutes developing the proofs and producing \texttt{answer.md}. It then completed the Lean formalization and committed the results in a further 22 minutes. These timings come from the retained platform event log and include tool latency and idle time.

\begin{table}[H]
  \caption{Summary of the autonomous theorem-generation run. Token counts are
  taken from the last record before the final answer and therefore exclude
  that turn.}
  \label{tab:autonomous-metrics}
  \centering
  \begin{tabular}{L{0.27\linewidth}L{0.63\linewidth}}
    \toprule
    Measure & Summary \\
    \midrule
    Messages
      & 272 intermediate assistant messages and one final answer. The context was compacted
        four times. \\
    Tool use
      & The log records 949 tool calls and 960 operations within those calls.
        The latter comprise 387 shell commands, 307 file edits, 205
        continuations of shell sessions, 55 goal checks, and six goal or plan
        updates. \\
    Token use
      & The platform reported about 142.4 million cumulative input tokens, of
        which 140.2 million were cached, and 0.64 million output tokens,
        including 0.32 million reasoning tokens. \\
    Research process
      & The agent registered thirteen approach families and used no public web
        search, external connector, or delegated model call. \\
    \bottomrule
  \end{tabular}
\end{table}

Input-token usage is cumulative across model calls, so retained context is counted repeatedly. Of the 142.4 million input tokens recorded, 98.5\% were retrieved from the prompt cache.

\paragraph{Exploring proof strategies.}
During the run, the agent tracked different approaches in \texttt{approach\_\allowbreak{}registry.md}. Table~\ref{tab:routes} groups these approaches by their goals and summarizes their outcomes.

\begin{table}[H]
  \caption{High-level summary of the proof families explored during the run.}
  \label{tab:routes}
  \centering
  \begin{tabular}{@{}L{0.20\linewidth}L{0.40\linewidth}L{\dimexpr0.40\linewidth-4\tabcolsep\relax}@{}}
    \toprule
    \textbf{Goal} & \textbf{Approach} & \textbf{Outcome} \\
    \midrule
    Reformulate head complexity
      & Replace the full matrix parameterization with a scalar description of
        each head's contribution.
      & Derived Theorem~\ref{thm:normal-form}. \\
    \addlinespace[0.45em]
    Prove upper bounds
      & Construct representations for functions determined by weighted sums
        and for symmetric functions.
      & Derived Theorems~\ref{thm:weighted-upper}
        and~\ref{thm:symmetric}, plus bounds not included in the paper. \\
    \addlinespace[0.45em]
    Prove lower bounds
      & Compare head complexity with threshold degree and study functions that
        require more heads.
      & Derived Theorem~\ref{thm:threshold} and additional lower-bound
        arguments. \\
    \addlinespace[0.45em]
    Evaluate candidate approaches
      & Check proposed identities and constructions on finite instances.
      & Rejected approaches that failed finite checks and set aside those that
        still required an unproved~lemma. \\
    \bottomrule
  \end{tabular}
\end{table}

The agent kept several distinct approaches under consideration rather than committing early to a single line of attack. Patterns observed in small cases were not treated as general results unless the remaining proof steps could be completed. When further exact checks produced counterexamples or showed that a proposed construction could not cover all cases, the corresponding approach was marked as blocked. This process prevented several plausible but unsupported generalizations from entering the final answer.

\subsection{Human interventions}
\label{subsec:human-interventions}

\paragraph{Theorem generation.}
The initial prompt specified the problem and the criteria for a complete answer. From the submission of this prompt to the delivery of \texttt{answer.md}, the agent received no further human input, mathematical lemmas, or corrections.

\paragraph{Lean formalization.}
After the theorem-generation run, the authors asked the agent to commit the artifacts and formalize the core results in Lean. The generated code captured the intended results but was not initially organized for readability or reuse. Human researchers selected the final theorem set, checked that the Lean statements matched the intended claims, and restructured the code to improve its naming, documentation, and use of the existing library.

\paragraph{Counterexample search.}
The decision to search for an explicit counterexample came from the authors. No such example had emerged during the autonomous theorem-generation runs. Given the one-line goal \emph{find an explicit constructive example when $\tdeg(f)$ is not equal to $\hstar(f)$}, the agent's first turn proceeded without further human input and produced a 128-bit example. In subsequent turns, the authors asked the agent to explain candidate examples, find progressively smaller examples with a Lean-friendly proof, and formalize the result in the shared repository. The search ultimately yielded the ten-bit function in Theorem~\ref{thm:separation}.

\subsection{Verification}

We rely on the Lean formalization to verify the reported theorems. The Lean kernel checks each proof, while a Lean expert checks that its formal statement faithfully captures the meaning, assumptions, and scope of the corresponding natural-language claim. This guards against hallucinated proofs. Future work could incorporate Lean's \texttt{comparator} tool to check untrusted, agent-generated proofs against human-reviewed theorem statements in a sandboxed environment \citep{leanfro2026comparator}.

\subsection{Significance of result}

The mathematical result treats the number of attention heads as an architectural resource and exactly characterizes the head count required to represent symmetric Boolean functions. It provides a tractable setting for studying architectural expressivity without modeling training dynamics. More generally, it establishes threshold degree as a lower bound on head complexity while showing that this lower bound need not be tight. These results therefore provide a mathematical reference point for asking whether a computation observed across several heads in a trained toy model is required by the architecture or reflects a particular solution found through training.

The broader program is to develop component-level results of this kind into mathematical abstractions for tracking features through attention, feed-forward networks, and multiple layers. Such abstractions could provide principled starting points for empirical studies and, eventually, inform feature-discovery tools for larger transformers. Many current feature-decomposition methods, including sparse autoencoders, learn features from model activations without constraining those features by the computations that the transformer's architecture can implement \citep{cunningham2023sparseautoencodershighlyinterpretable,shafran2026from}. 

An architecture-aware probe could instead ask how many attention heads are required to construct a given feature in a single attention-layer update. The linear-fractional normal form in Theorem~\ref{thm:normal-form}, which characterizes the contribution realizable by one attention head, provides a theoretical starting point for developing such a probe. The present result is an initial step: it gives exact statements for a one-layer attention only transformer rather than demonstrating transfer to trained models at scale.

Together, the mathematical results and their Lean formalization give other mechanistic interpretability researchers a concrete artifact to reuse and extend. Although the current development is limited to the one-layer Boolean setting studied here, it lays the groundwork for a reusable formal environment in which related questions about transformer components can be stated and verified.

\subsection{Interpretability}

During the autonomous theorem-generation run, the agent maintained a registry of approaches and their status, and executable scripts for computational checks. The platform recorded prompts, assistant messages and tool calls that can be useful when we want to audit the autoresearch run. Together, these records allow the research trajectory to be reconstructed, including why some approaches were completed or abandoned.

\subsection{Limitations}

\paragraph{Auditability of autonomous research.}
The volume of reasoning and results produced during long autonomous runs makes their progress difficult to inspect and steer. Determining which approaches advanced, failed, or were abandoned requires substantial review. Tools that summarize research branches and track the status of claims could reduce this burden and help researchers identify where human guidance would be most useful.

\paragraph{Repetition and premature claims of progress.}
Extended runs can produce repetitive updates and apparent progress that does not survive scrutiny. The retained rollout contains prolonged monitoring of stalled searches, an incorrect explanation repeated before correction, and a proposed new bound later removed because an existing bound was stronger. These episodes added work without a corresponding improvement in the final result. A monitor that tracks new evidence, flags repetitive or unsupported claims, and recommends redirection or stopping could help guide the search. Whether such monitoring improves research outcomes remains untested.

Together, these limitations motivate treating autoresearch itself as an object of scientific study. Controlled comparisons of prompts, tools, monitoring, human steering, and stopping rules are needed to determine which workflows produce reliable results and how much computational and human effort they require.

\section{Broader Impact}

\paragraph{Societal benefits and risks.}
This work contributes to a longer-term effort to make transformer behavior more understandable and amenable to oversight. Precise results about architectural expressivity can provide foundations for interpretability methods, which may eventually help researchers identify and investigate undesirable model behavior by designing theoretically grounded feature-decomposition tools. These benefits remain prospective: our theorems concern exact representation in a one-layer attention only model and do not establish safety guarantees for deployed systems. Furthermore, the research process in this work serves as a proof of concept for mechanistic interpretability researchers who want to pursue mathematically demanding questions but lack ready access to collaborators with expertise in mathematical research or formal theorem proving. 

\paragraph{Scientific judgment.}
Autonomous theorem discovery leaves a central role for human scientific judgment, both in asking informative questions and in assessing the resulting claims. Since the overarching question is broad and has no obvious route to a complete answer, researchers must choose tractable idealizations that isolate meaningful aspects of transformer computation that can inform empirical interpretability. While the state of the art agents are good at solving well defined problems, it remains for the researcher to ask the high level questions and guide the autoresearch runs.

\paragraph{Understanding failures and steering autoresearch.}
We need to understand its failure modes and how to steer it, especially for open-ended theoretical questions, where no obvious scalar metric captures scientific progress. We propose adapting Inspect~\citep{aisi2024inspect} and Inspect Scout~\citep{dubois2026loganalysis} to audit research runs through their logs. A monitoring agent built on this infrastructure could track verified intermediate results and flag repeated attempts that do not contribute to research progress.

\appendix
\section{Reproducibility Statement}
All results in this work are accompanied by a public repository available at \url{https://github.com/VerifiedMechanisms/automlr}. The natural-language theorem statements and proofs are collected in the top-level \texttt{theorems/} directory. The corresponding machine-checked Lean~4 development is contained in \texttt{formalization/}. The theorem-generation and counterexample-search records are released in \texttt{system-design/rollouts/}. Build and proof-integrity checks are documented in \texttt{formalization/README.md} and \texttt{formalization/BUILDING.md}.

The repository is self contained and no external datasets are required to reproduce the results. The authors did the 9.5 hour autoresearch run using a ChatGPT Pro plan. The machine-checked results in this work are done using Lean~4.31.0 and Mathlib~4.31.0. Refer to the \texttt{README.md} in the accompanying repository for detailed instructions. 

\section{Disclosure Statement}

\subsection*{Agent(s) / model(s) used}

All theorems and their accompanying Lean formalizations were generated with \gpt{} in Codex Desktop. Codex also produced the initial draft of the technical paper and assisted with sentence-level revisions. The human authors refined the sections presenting the autonomous research results and wrote the system-design section.

\begin{itemize}[leftmargin=1.8em,nosep]
  \item[$\boxed{\checkmark}$] A human author has curated this work and verified every claim, figure, methodology detail, and reference.
  \item[$\boxed{\checkmark}$]  A human author takes full accountability for the submission as a human-authored, human-verified artifact.
\end{itemize}

\begin{ack}
This work was supported in part by a Rapid Grant from BlueDot Impact. The grant funded AI tool subscriptions used during the Lean formalization. Karthik Viswanathan was funded by AI Safety Support (AISS) during this work. Karthik Viswanathan worked on this project as part of the MATS program, and thanks Nayara Fonseca, Paul Riechers, Adam Shai, and Dmitry Manning-Coe for productive discussions and support through the research process. We thank \href{https://talmenta.org}{Talmenta}, Jasper van de Kreeke, and Fahad Sarwar for their help with research management.
\end{ack}

\renewcommand{\bibsection}{\section{References}}
\bibliographystyle{plainnat}
\bibliography{references}

\section{Relation to Concurrent Work}
\label{app:hsu-comparison}

\paragraph{Overlapping results with Hsu and Xu.}
In concurrent work, \citet{hsu2026attentionbased} prove that an $H$-head attention layer with a degree-$D$ polynomial-threshold readout representing $f$ must satisfy $DH\geq\tdeg(f)$, and that this lower bound is exact for symmetric functions. At $D=1$, their result agrees with our symmetric characterization in Theorem~\ref{thm:symmetric}, giving the same minimum head count for nonconstant symmetric functions. We developed the symmetric characterization independently of the publicly released Hsu--Xu manuscript.

\paragraph{Differences from Hsu and Xu.}
Their model omits positional embeddings, whereas our fixed-query, linear-readout model allows arbitrary additive positional embeddings shared across heads. Theorem~\ref{thm:symmetric} therefore shows that positional information does not lower the required head count for symmetric functions. Beyond the symmetric setting, we additionally derive an exact linear-fractional normal form, prove the universal bound $\hstar(f)\leq 2^n-1$, and exhibit a nonsymmetric function with $\tdeg(f)=2<3\leq\hstar(f)$, showing that threshold degree does not characterize head complexity. Additionally, we also make publicly available a Lean~4 formalization of the attention model and the main results, including the symmetric characterization, the linear-fractional normal form, and the nonsymmetric separation.

\paragraph{Head complexity bounds of Rajaraman et al.}
\citet{rajaraman2026headcomplexity} prove that $n$-bit parity needs exactly $n$ heads, agreeing with Corollary~\ref{cor:parity}. Their lower bounds permit arbitrary embeddings of token-position pairs, while their constructions use additive embeddings. Their Lemma~8 also obtains a polynomial sign representation of degree at most the head count by clearing positive denominators, matching the implication of our Theorem~\ref{thm:threshold} in a more general embedding model. They further establish a $2^n$ universal upper bound, an $\Omega(2^n/n^2)$ lower bound for almost all Boolean functions, and dimension and precision compactness bounds.

\section{Model Details}
\label{app:model-details}

This section gives a self-contained specification of the model underlying Equations~\eqref{eq:output} and~\eqref{eq:readout}. Fix an input length $n$ and a number of heads $H$. The model receives an $n$-bit input, applies a single multi-head self-attention update, and classifies the resulting representation at a dedicated query position. Figure~\ref{fig:model} gives an overview of this computation.

\paragraph{Input sequence and representations.}
For $\mathbf{x}=(x_1,\ldots,x_n)\in\bits^n$, append a dedicated query token $q$ to form the sequence $(x_1,\ldots,x_n,q)$. The query token is a distinguished vocabulary element, for which the symbol $=$ is one possible choice. Let $\mathcal{I}:=[n]\cup\{q\}$ denote the set of sequence positions.

Choose a model dimension $d_{\mathrm{model}}$, token embeddings $\mathbf{e}_0,\mathbf{e}_1,\mathbf{e}_q\in\mathbb{R}^{d_{\mathrm{model}}}$, and positional embeddings $\mathbf{p}_1,\ldots,\mathbf{p}_n,\mathbf{p}_q\in\mathbb{R}^{d_{\mathrm{model}}}$. Following the standard additive positional-encoding convention, the representation at each position is the sum of its token embedding and its positional embedding:
\begin{equation}
\mathbf{u}_j(\mathbf{x})=
\begin{cases}
\mathbf{e}_{x_j}+\mathbf{p}_j, & j\in[n],\\
\mathbf{e}_q+\mathbf{p}_q, & j=q.
\end{cases}
\label{eq:app-input-representations}
\end{equation}
In each case, the first term identifies the token occupying the position and the second encodes the position itself. Since the token at the query position is always $q$, $\mathbf{u}_q$ is independent of $\mathbf{x}$. These token and positional embeddings are shared across heads; each head operates on the same embedded sequence $\{\mathbf{u}_j(\mathbf{x})\}_{j\in\mathcal I}$.

\paragraph{Single-head attention update.}
Choose a head dimension $d_{\mathrm{head}}$. Each head $h\in[H]$ has its own query, key, and value projections
\begin{equation}
W_Q^{(h)},W_K^{(h)},W_V^{(h)}
\in\mathbb{R}^{d_{\mathrm{head}}\times d_{\mathrm{model}}}
\end{equation}
and output projection $W_O^{(h)}\in\mathbb{R}^{d_{\mathrm{model}}\times d_{\mathrm{head}}}$. At the query position, head $h$ assigns position $j\in\mathcal I$ the logit and normalized attention weight
\begin{align}
\ell_j^{(h)}(\mathbf{x})
&=\left\langle W_K^{(h)}\mathbf{u}_j(\mathbf{x}),
W_Q^{(h)}\mathbf{u}_q\right\rangle,
\label{eq:app-attention-logit}\\
a_j^{(h)}(\mathbf{x})
&=\frac{\exp\!\left(\ell_j^{(h)}(\mathbf{x})\right)}
{\sum_{r\in\mathcal I}\exp\!\left(\ell_r^{(h)}(\mathbf{x})\right)}.
\label{eq:app-attention-weight}
\end{align}
Its contribution to the query position is
\begin{equation}
\mathbf{y}_q^{(h)}(\mathbf{x})
=W_O^{(h)}\sum_{j\in\mathcal I}
a_j^{(h)}(\mathbf{x})W_V^{(h)}\mathbf{u}_j(\mathbf{x})
\in\mathbb{R}^{d_{\mathrm{model}}}.
\label{eq:app-head-output}
\end{equation}
Only the output at $q$ is used for readout; outputs at other positions are not analyzed in this work.

\paragraph{Multi-head output and readout.}
The projected head outputs are added to the residual stream at the query position:
\begin{equation}
\mathbf{h}_q(\mathbf{x})
=\mathbf{u}_q+\sum_{h=1}^{H}\mathbf{y}_q^{(h)}(\mathbf{x}).
\label{eq:app-multihead-output}
\end{equation}
Given a readout vector $\mathbf{w}\in\mathbb{R}^{d_{\mathrm{model}}}$ and threshold $\tau\in\mathbb{R}$, an $H$-head configuration exactly represents $f:\bits^n\to\bits$ if
\begin{equation}
\forall\mathbf{x}\in\bits^n,\qquad
\bigl[f(\mathbf{x})=1
\quad\Longleftrightarrow\quad
\left\langle\mathbf{w},\mathbf{h}_q(\mathbf{x})\right\rangle>\tau\bigr].
\label{eq:app-linear-readout}
\end{equation}
The head complexity $\hstar(f)$ is the minimum $H\in\mathbb{Z}_{\geq0}$ over all choices of dimensions, embeddings, projections, and readout parameters that exactly represent $f$.

\paragraph{Conventions and scope.}
All parameters are real-valued, and $d_{\mathrm{model}}$ and $d_{\mathrm{head}}$ are unrestricted. The usual factor $1/\sqrt{d_{\mathrm{head}}}$ in scaled dot-product attention is omitted because it can be absorbed into the query or key projections. Likewise, $W_O^{(h)}$ can be composed with $W_V^{(h)}$; we retain the output projections to distinguish the model and head dimensions. The residual $\mathbf{u}_q$ is independent of $\mathbf{x}$ and can therefore be absorbed into the readout threshold. Unrestricted width allows the shared embeddings to allocate orthogonal coordinate blocks selected by different heads.

The query token occurs last, so full attention and a standard causal mask expose it to the same set of positions. The architecture contains no MLP, layer normalization, dropout, or additional layers.

\section{Proofs of the Main Results}
\label{app:proofs}

\subsection{Linear-Fractional Normal Form}
\label{app:proof-normal-form}

We prove that the admissibility conditions in Definition~\ref{def:atom} characterize exactly the coefficients obtained from the one-head parameterization. We then establish both directions of the scalar-score representation in Theorem~\ref{thm:normal-form}, including simultaneous realization of all atoms with shared embeddings.

\paragraph{Parameterization and affine coefficients.}
For $\gamma,\alpha,\rho_1,\ldots,\rho_n>0$ and $\eta,\beta,m_1,\ldots,m_n\in\mathbb{R}$, define
\begin{equation}
\phi(\mathbf{x})=
\frac{\eta+\sum_{i=1}^{n}\rho_i\alpha^{x_i}(m_i+\beta x_i)}
{\gamma+\sum_{i=1}^{n}\rho_i\alpha^{x_i}},
\qquad \mathbf{x}\in\bits^n.
\label{eq:atom-parameterization}
\end{equation}
For Boolean inputs, $\alpha^{x_i}=1+(\alpha-1)x_i$ and $x_i^2=x_i$. Expanding the numerator and denominator therefore gives Equation~\eqref{eq:frac-atom}, with
\begin{equation}
\begin{aligned}
A_i&=\rho_i\bigl((\alpha-1)m_i+\alpha\beta\bigr),
& b&=\eta+\sum_i\rho_i m_i,\\
C_i&=\rho_i(\alpha-1),
& d&=\gamma+\sum_i\rho_i.
\end{aligned}
\label{eq:affine-atom-coefficients}
\end{equation}

\paragraph{Exact coefficient constraints.}
A tuple $(\mathbf{A},b,\mathbf{C},d)$ is obtained from Equation~\eqref{eq:affine-atom-coefficients} if and only if it is admissible in the sense of Definition~\ref{def:atom}. We prove necessity directly from the coefficient formulas, then give explicit parameters for every admissible tuple.

Condition~(i) of Definition~\ref{def:atom} is equivalent to $d+\sum_i\min\{0,C_i\}>0$. Under condition~(ii), it therefore reduces to $d>0$ when the denominator slopes are positive or all zero, and to $d+\sum_i C_i>0$ when they are negative. A mixture of zero and nonzero denominator slopes is excluded.

\paragraph{Necessity.}
The denominator in Equation~\eqref{eq:atom-parameterization} is strictly positive on the Boolean cube. Since $C_i=\rho_i(\alpha-1)$ and every $\rho_i>0$, its slopes have a common strict sign when $\alpha\ne1$. When $\alpha=1$, all $C_i=0$ and $A_i=\rho_i\beta$, giving the stated restriction on the numerator slopes. The case $n=0$ has no slopes and only requires $d>0$.

\paragraph{Sufficiency.}
Fix an admissible tuple. Evaluating its denominator at $\mathbf{x}=\mathbf{0}$ gives $d>0$. If $\mathbf{C}\ne\mathbf{0}$, set $S=\sum_i|C_i|$ and choose
\begin{equation}
\alpha=\begin{cases}
2+S/d, & C_i>0\text{ for all }i,\\
(d-S)/(2d), & C_i<0\text{ for all }i.
\end{cases}
\end{equation}
In the negative case, denominator positivity at $\mathbf{x}=\mathbf{1}$ gives $S<d$. In both cases, $\alpha>0$, and the assignments
\begin{equation}
\begin{aligned}
\rho_i&=\frac{C_i}{\alpha-1},
& \gamma&=d-\sum_i\rho_i,\\
\beta&=0,
& m_i&=\frac{A_i}{C_i},
\qquad \eta=b-\sum_i\rho_i m_i
\end{aligned}
\end{equation}
have $\rho_i>0$ and $\sum_i\rho_i<d$, hence $\gamma>0$. Substitution into Equation~\eqref{eq:affine-atom-coefficients} recovers the prescribed tuple exactly.

If $\mathbf{C}=\mathbf{0}$ and $\mathbf{A}\ne\mathbf{0}$, let $\sigma\in\{-1,1\}$ be the common sign of the $A_i$ and put $q=1+\sum_i|A_i|/d$. Set
\begin{equation}
\alpha=1,\quad \beta=\sigma q,\quad
\rho_i=|A_i|/q,\quad \gamma=d-\sum_i\rho_i,\quad
m_i=0,\quad \eta=b.
\end{equation}
Then $\rho_i>0$, $\gamma>0$, and $\rho_i\beta=A_i$. If $\mathbf{A}=\mathbf{C}=\mathbf{0}$, take $\alpha=1$, $\beta=m_i=0$, $\eta=b$, and $\rho_i=\gamma=d/(n+1)$. This also covers $n=0$. Thus every admissible coefficient tuple comes from the parameterization.

\paragraph{Coefficient representations.}
Admissibility concerns the chosen coefficients. A function is an atom whenever it has \emph{some} admissible representation, even if another representation of the same function fails the conditions. For example, on $\bits^2$,
\begin{equation*}
x_1-x_2=\frac{2x_1-2x_2}{1+x_1+x_2}.
\end{equation*}
The constant-denominator representation on the left has mixed numerator slopes and fails condition~(iii) of Definition~\ref{def:atom}, while the representation on the right is admissible. This identity, like the affine expansion of each head, uses the Boolean-input assumption.

\paragraph{Decomposing the readout score.}
Fix an $H$-head model with readout $(\mathbf{w},\tau)$, and let $\mathbf{y}_q^{(h)}(\mathbf{x})$ denote the projected output of head $h$, that is, the $h$-th summand in Equation~\eqref{eq:output}. Define its scalar contribution after readout by
\begin{equation}
g^{(h)}(\mathbf{x})
:=\left\langle\mathbf{w},\mathbf{y}_q^{(h)}(\mathbf{x})\right\rangle,
\qquad h\in[H],
\label{eq:normal-head-contribution}
\end{equation}
Taking the inner product of Equation~\eqref{eq:output} with $\mathbf{w}$ then gives
\begin{equation}
\left\langle\mathbf{w},\mathbf{h}_q(\mathbf{x})\right\rangle
=\left\langle\mathbf{w},\mathbf{u}_q\right\rangle
+\sum_{h=1}^{H}g^{(h)}(\mathbf{x}),
\label{eq:normal-score-decomposition}
\end{equation}
Subtracting the threshold $\tau$ gives the input-independent constant $c=\langle\mathbf{w},\mathbf{u}_q\rangle-\tau$ in Equation~\eqref{eq:score-normal-form}. Thus, the forward direction of Theorem~\ref{thm:normal-form} reduces to showing that each $g^{(h)}$ is a linear-fractional atom.

\paragraph{From a head to an atom.}
\label{app:from-head-to-atom}
For a head $h$, define
\begin{equation}
\mathbf{q}^{(h)}=W_Q^{(h)}\mathbf{u}_q,
\qquad
\Delta\mathbf{e}=\mathbf{e}_1-\mathbf{e}_0.
\end{equation}
The query vector $\mathbf{q}^{(h)}$ is independent of the input. Since $x_i\in\{0,1\}$, the representation at input position $i$ can be written as
\begin{equation}
\mathbf{u}_i(\mathbf{x})
=\mathbf{e}_{x_i}+\mathbf{p}_i
=\mathbf{e}_0+\mathbf{p}_i+x_i\Delta\mathbf{e}.
\label{eq:normal-input-decomposition}
\end{equation}
Let $\ell_j^{(h)}(\mathbf{x})$ denote the attention logit at position $j$, and let $s_j^{(h)}(\mathbf{x})$ denote its scalar value after the value projection, output projection, and readout:
\begin{equation}
\ell_j^{(h)}(\mathbf{x})
=\left\langle W_K^{(h)}\mathbf{u}_j(\mathbf{x}),\mathbf{q}^{(h)}\right\rangle,
\qquad
s_j^{(h)}(\mathbf{x})
=\left\langle\mathbf{w},W_O^{(h)}W_V^{(h)}\mathbf{u}_j(\mathbf{x})\right\rangle.
\end{equation}
Substituting Equation~\eqref{eq:normal-input-decomposition} into the logit gives
\begin{align}
\ell_i^{(h)}(\mathbf{x})
&=\left\langle W_K^{(h)}(\mathbf{e}_0+\mathbf{p}_i),\mathbf{q}^{(h)}\right\rangle
+x_i\left\langle W_K^{(h)}\Delta\mathbf{e},\mathbf{q}^{(h)}\right\rangle,
\label{eq:normal-logit-expansion}\\
\exp\!\left(\ell_i^{(h)}(\mathbf{x})\right)
&=\rho_i^{(h)}\bigl(\alpha^{(h)}\bigr)^{x_i},
\label{eq:normal-logit-factorization}
\end{align}
where
\begin{equation}
\rho_i^{(h)}
=\exp\!\left(\left\langle W_K^{(h)}(\mathbf{e}_0+\mathbf{p}_i),\mathbf{q}^{(h)}\right\rangle\right),
\qquad
\alpha^{(h)}
=\exp\!\left(\left\langle W_K^{(h)}\Delta\mathbf{e},\mathbf{q}^{(h)}\right\rangle\right).
\end{equation}
Likewise, substituting Equation~\eqref{eq:normal-input-decomposition} into the scalar value gives
\begin{align}
s_i^{(h)}(\mathbf{x})
&=m_i^{(h)}+\beta^{(h)}x_i,
\label{eq:normal-value-factorization}\\
m_i^{(h)}
&=\left\langle\mathbf{w},W_O^{(h)}W_V^{(h)}(\mathbf{e}_0+\mathbf{p}_i)\right\rangle,
\qquad
\beta^{(h)}
=\left\langle\mathbf{w},W_O^{(h)}W_V^{(h)}\Delta\mathbf{e}\right\rangle.
\end{align}
Thus $\rho_i^{(h)}$ and $m_i^{(h)}$ are the unnormalized attention weight and the value of position $i$ when $x_i=0$, while $\alpha^{(h)}$ is the factor gained and $\beta^{(h)}$ the shift incurred when $x_i=1$. The latter two depend only on $\Delta\mathbf{e}$, so they are the same at every position. At the query position, both quantities are input-independent. Set
\begin{equation}
\gamma^{(h)}
=\exp\!\left(\left\langle W_K^{(h)}\mathbf{u}_q,\mathbf{q}^{(h)}\right\rangle\right),
\qquad
\eta^{(h)}
=\gamma^{(h)}
\left\langle\mathbf{w},W_O^{(h)}W_V^{(h)}\mathbf{u}_q\right\rangle.
\label{eq:normal-query-parameters}
\end{equation}
The scalar contribution of head $h$ to the readout score is therefore
\begin{align}
g^{(h)}(\mathbf{x})
&=\frac{\sum_{j\in\mathcal{I}}
\exp\!\left(\ell_j^{(h)}(\mathbf{x})\right)s_j^{(h)}(\mathbf{x})}
{\sum_{j\in\mathcal{I}}\exp\!\left(\ell_j^{(h)}(\mathbf{x})\right)}
\notag\\
&=\frac{\eta^{(h)}+\sum_{i=1}^{n}\rho_i^{(h)}
\bigl(\alpha^{(h)}\bigr)^{x_i}
\bigl(m_i^{(h)}+\beta^{(h)}x_i\bigr)}
{\gamma^{(h)}+\sum_{i=1}^{n}\rho_i^{(h)}
\bigl(\alpha^{(h)}\bigr)^{x_i}}.
\label{eq:normal-head-calculation}
\end{align}
The exponential definitions give $\rho_i^{(h)},\alpha^{(h)},\gamma^{(h)}>0$, so Equation~\eqref{eq:normal-head-calculation} has the parameterization in Equation~\eqref{eq:atom-parameterization}. Applying Equation~\eqref{eq:affine-atom-coefficients} to head $h$ yields the affine ratio in Equation~\eqref{eq:frac-atom}, with admissible coefficients. Hence $g^{(h)}$ is an atom in the sense of Definition~\ref{def:atom}. By Equation~\eqref{eq:normal-score-decomposition}, the complete pre-threshold score is
\begin{equation}
\left\langle\mathbf{w},\mathbf{h}_q(\mathbf{x})\right\rangle-\tau
=\underbrace{\left\langle\mathbf{w},\mathbf{u}_q\right\rangle-\tau}_{c}
+\sum_{h=1}^{H}g^{(h)}(\mathbf{x}).
\end{equation}
Thus every $H$-head representation gives an $H$-atom representation, proving $L_{\mathrm{frac}}(f)\leq\hstar(f)$.

\paragraph{From atoms to heads.}
Conversely, consider a score $c+\sum_{h=1}^{H}\phi_h(\mathbf{x})$ with atoms as in Definition~\ref{def:atom}. For each atom, choose parameters in Equation~\eqref{eq:atom-parameterization} that produce its affine coefficients, and index these parameters by $h$. The coefficient characterization above guarantees such a choice. If $H=0$, take $\mathbf{w}=0$ and $\tau=-c$ to realize the constant score. For $H\geq1$, choose $d_{\mathrm{model}}\geq4H$ and orthonormal vectors $\mathbf{q}^{(h)},\mathbf{k}^{(h)},\mathbf{v}^{(h)},\mathbf{o}^{(h)}\in\mathbb{R}^{d_{\mathrm{model}}}$ for each $h$. Let $d_{\mathrm{head}}\geq1$ be arbitrary, and fix a unit vector $\mathbf{r}\in\mathbb{R}^{d_{\mathrm{head}}}$. Choose the embeddings
\begin{align}
\mathbf{e}_0&=0,
&
\mathbf{e}_1&=\sum_h\left((\log\alpha^{(h)})\mathbf{k}^{(h)}
+\beta^{(h)}\mathbf{v}^{(h)}\right),
\label{eq:normal-token-embeddings}
\end{align}
\begin{equation}
\begin{aligned}
\mathbf{e}_q&=\sum_h\left(\mathbf{q}^{(h)}
+(\log\gamma^{(h)})\mathbf{k}^{(h)}
+\frac{\eta^{(h)}}{\gamma^{(h)}}\mathbf{v}^{(h)}\right),
\\
\mathbf{p}_i&=\sum_h\left((\log\rho_i^{(h)})\mathbf{k}^{(h)}
+m_i^{(h)}\mathbf{v}^{(h)}\right),
\qquad \mathbf{p}_q=0.
\end{aligned}
\label{eq:normal-position-embeddings}
\end{equation}
For each head $h$, define the linear maps on these directions by
\begin{equation}
W_Q^{(h)}\mathbf{q}^{(h)}=\mathbf{r},
\qquad
W_K^{(h)}\mathbf{k}^{(h)}=\mathbf{r},
\qquad
W_V^{(h)}\mathbf{v}^{(h)}=\mathbf{r},
\qquad
W_O^{(h)}\mathbf{r}=\mathbf{o}^{(h)},
\label{eq:normal-projection-construction}
\end{equation}
extending each map by zero on the orthogonal complement of its displayed input direction. The logits at input position $i$ and the query position are
\begin{equation}
\log\rho_i^{(h)}+x_i\log\alpha^{(h)}
\qquad\text{and}\qquad
\log\gamma^{(h)},
\end{equation}
and the corresponding scalar values in the $\mathbf{o}^{(h)}$ direction are $m_i^{(h)}+\beta^{(h)}x_i$ and $\eta^{(h)}/\gamma^{(h)}$. Hence head $h$ writes $\phi_h(\mathbf{x})\mathbf{o}^{(h)}$. Choose $\mathbf{w}=\sum_h\mathbf{o}^{(h)}$ and $\tau=-c$. The query residual is orthogonal to the readout, so the resulting scalar score is exactly $c+\sum_h\phi_h(\mathbf{x})$ on every Boolean input. This proves the reverse direction of Equation~\eqref{eq:score-normal-form}. Thresholding the score and minimizing over $H$ yields $\hstar(f)=L_{\mathrm{frac}}(f)$.

\subsection{One-Head Characterization}
\label{app:proof-one-head}

We prove Corollary~\ref{cor:one-head}.

\paragraph{Zero heads.}
When $H=0$, Equation~\eqref{eq:output} reduces to $\mathbf{h}_q(\mathbf{x})=\mathbf{u}_q$, which is independent of $\mathbf{x}$. Every zero-head classifier is therefore constant. Conversely, either constant function can be realized by placing the readout threshold above or below the fixed score $\langle\mathbf{w},\mathbf{u}_q\rangle$. Hence $\hstar(f)=0$ if and only if $f$ is constant.

\paragraph{At most one head implies linear threshold.}
Suppose $\hstar(f)\leq1$. The zero-head case is covered above. Otherwise, Theorem~\ref{thm:normal-form} gives a constant $c$ and one atom $\phi=N/D$ such that $f(\mathbf{x})=1$ exactly when $c+\phi(\mathbf{x})>0$. The numerator $N$ and denominator $D$ are affine on the Boolean cube, and $D(\mathbf{x})>0$. Therefore
\begin{equation}
c+\phi(\mathbf{x})
=\frac{cD(\mathbf{x})+N(\mathbf{x})}{D(\mathbf{x})}
\end{equation}
has the same sign as the affine function $cD+N$. Thus $f$ is a linear threshold function.

\paragraph{Linear threshold implies at most one head.}
Suppose $f(\mathbf{x})=1$ exactly when $\theta_0+\sum_{i=1}^{n}\theta_ix_i>0$. Consider
\begin{equation}
\phi(\mathbf{x})=\frac{\theta_0+\sum_{i=1}^{n}\theta_ix_i}{1+\sum_{i=1}^{n}x_i}.
\end{equation}
Its denominator is strictly positive and has all slopes equal to one. Hence its coefficients are admissible by Definition~\ref{def:atom}, and it is an atom with the same sign as the given affine threshold function. This also covers $n=0$, when both slope vectors are empty. Taking $c=0$ in Definition~\ref{def:lfrac} yields $L_{\mathrm{frac}}(f)\leq1$, and Theorem~\ref{thm:normal-form} then gives $\hstar(f)\leq1$.

\subsection{Threshold-Degree Lower Bound}
\label{app:proof-threshold}

We prove Theorem~\ref{thm:threshold}. Equation~\eqref{eq:frac-atom} writes each atom as a ratio of affine functions with a denominator that is strictly positive on the Boolean cube. By Theorem~\ref{thm:normal-form}, after moving the readout threshold into a constant, an $H$-head classifier therefore has a scalar score
\begin{equation}
S(\mathbf{x})=c+\sum_{h=1}^{H}\frac{N_h(\mathbf{x})}{D_h(\mathbf{x})},
\qquad D_h(\mathbf{x})>0,
\label{eq:appendix-rational-score}
\end{equation}
where $N_h$ and $D_h$ are affine. Since the domain is finite, the threshold can be perturbed so that $S$ is strictly positive on inputs labelled one and strictly negative on inputs labelled zero. Multiplication by the positive common denominator preserves this sign pattern and gives
\begin{equation}
P(\mathbf{x})=
c\prod_{h=1}^{H}D_h(\mathbf{x})+
\sum_{h=1}^{H}N_h(\mathbf{x})\prod_{g\neq h}D_g(\mathbf{x}).
\label{eq:appendix-cleared-score}
\end{equation}
Every summand has degree at most $H$, so $P$ is a degree-at-most-$H$ sign representation. Minimizing over $H$ proves $\tdeg(f)\leq\hstar(f)$.

\subsection{Strict Separation from Threshold Degree}
\label{app:proof-separation}

We prove Theorem~\ref{thm:separation} for the function in Equation~\eqref{eq:f10}. For its Boolean blocks $\mathbf{a},\mathbf{b}\in\bits^5$, introduce the sign coordinates
\begin{equation}
s_i=1-2a_i,
\qquad
t_i=1-2b_i,
\qquad i\in[5],
\label{eq:f10-sign-coordinates}
\end{equation}
so $\mathbf{s},\mathbf{t}\in\{-1,+1\}^5$. Define the corresponding quadratic form on the sign cube by
\begin{equation}
\widetilde Q(\mathbf{s},\mathbf{t})=
\left(\sum_{i=1}^{5}s_i\right)\left(\sum_{j=1}^{5}t_j\right)
-3\sum_{i=1}^{5}s_it_i.
\label{eq:f10-signed-form}
\end{equation}
Equation~\eqref{eq:f10} then gives $Q(\mathbf{a},\mathbf{b})=\widetilde Q(\mathbf{s},\mathbf{t})$. The coordinate change is an invertible affine map between the Boolean and sign cubes. Substitution in either direction does not increase polynomial degree, so threshold degree is unchanged; affine forms also remain affine.

Let $A$ and $B$ be the numbers of negative entries in $\mathbf{s}$ and $\mathbf{t}$, and let $C$ count coordinates at which both are negative. Direct substitution gives
\begin{equation}
\widetilde Q(\mathbf{s},\mathbf{t})=10+4(-A-B+AB-3C)\equiv2\pmod 4.
\label{eq:f10-mod-four}
\end{equation}
Thus $\widetilde Q$ never vanishes, so the polynomial $Q$ in Equation~\eqref{eq:f10} is a strict degree-two sign representation of $f_{10}$. On the two-dimensional face of the sign cube
\begin{equation}
\mathbf{s}=(r,1,1,-1,-1),\qquad
\mathbf{t}=(w,1,-1,1,-1),\qquad r,w\in\{-1,+1\},
\end{equation}
we have $\widetilde Q(\mathbf{s},\mathbf{t})=-2rw$. Under the inverse coordinate change, this is a two-dimensional face of the original Boolean cube whose sign pattern is XOR, so it cannot be sign-represented by an affine function. Consequently, $\tdeg(f_{10})=2$.

It remains to rule out two heads. If two heads represented $f_{10}$, Equation~\eqref{eq:appendix-cleared-score} would give a strict sign-representing polynomial
\begin{equation}
P=cD_1D_2+N_1D_2+N_2D_1=(cD_1+N_1)D_2+N_2D_1.
\label{eq:two-head-regrouping}
\end{equation}
Since $N_1,N_2,D_1,D_2$ are affine, this is a sum of two products of affine forms. Pull this polynomial back to the sign cube by setting
\begin{equation}
P^{\pm}(\mathbf{s},\mathbf{t})=
P\left(\frac{\mathbf{1}-\mathbf{s}}{2},
       \frac{\mathbf{1}-\mathbf{t}}{2}\right),
\end{equation}
where $\mathbf{1}$ denotes the all-ones vector. Because the coordinate change is affine, the regrouping retains the form
\begin{equation}
P^{\pm}=L_1R_1+L_2R_2,
\label{eq:two-products}
\end{equation}
where all four factors are affine in $(\mathbf{s},\mathbf{t})$, and $P^{\pm}$ has the sign pattern of $\widetilde Q$. For a function $g$ on the two signed blocks, define its mixed antipodal difference
\begin{equation}
\Delta g(\mathbf{s},\mathbf{t})=
g(\mathbf{s},\mathbf{t})-g(\mathbf{s},-\mathbf{t})
-g(-\mathbf{s},\mathbf{t})+g(-\mathbf{s},-\mathbf{t}).
\end{equation}
Writing an affine form as $L(\mathbf{s},\mathbf{t})=\ell_0+\mathbf{c}_L\cdot\mathbf{s}+\mathbf{d}_L\cdot\mathbf{t}$ gives
\begin{equation}
\Delta(LR)(\mathbf{s},\mathbf{t})=
4\bigl[(\mathbf{c}_L\cdot\mathbf{s})(\mathbf{d}_R\cdot\mathbf{t})
+(\mathbf{c}_R\cdot\mathbf{s})(\mathbf{d}_L\cdot\mathbf{t})\bigr].
\label{eq:mixed-product}
\end{equation}
Each product $L_hR_h$ therefore contributes two outer products, $4\mathbf{c}_{L_h}\mathbf{d}_{R_h}^{\top}$ and $4\mathbf{c}_{R_h}\mathbf{d}_{L_h}^{\top}$, to the mixed coefficient matrix. Each has rank at most one. Thus $\Delta P^{\pm}(\mathbf{s},\mathbf{t})=\mathbf{s}^{\top}K\mathbf{t}$ for a matrix $K$ with $\operatorname{rank}(K)\leq4$.

For $j\in[5]$, let $\mathbf{z}^{(j)}$ have entry $-1$ in coordinate $j$ and $+1$ elsewhere. Equation~\eqref{eq:f10-signed-form} gives
\begin{equation}
\widetilde Q(\mathbf{s},\mathbf{z}^{(j)})=6s_j,
\end{equation}
and negating either whole block reverses its sign. Set $Z=[\mathbf{z}^{(1)}\ \cdots\ \mathbf{z}^{(5)}]$ and $M=KZ$. The $j$th column of $M$ is the coefficient vector of the slice $\Delta P^{\pm}(\mathbf{s},\mathbf{z}^{(j)})=\sum_i M_{ij}s_i$, and $\operatorname{rank}(M)\leq\operatorname{rank}(K)\leq4$. Since $P^{\pm}$ has the sign pattern of $\widetilde Q$, all four terms in this difference have sign $s_j$, giving
\begin{equation}
s_j\sum_iM_{ij}s_i>0
\qquad\text{for every }\mathbf{s}\in\{-1,+1\}^5.
\end{equation}
Fixing $s_j=1$ and choosing every other sign to oppose $M_{ij}$ yields
\begin{equation}
M_{jj}>\sum_{i\neq j}|M_{ij}|.
\end{equation}
Thus $M$ is strictly column diagonally dominant and therefore invertible, so it has rank five. This contradicts the bound $\operatorname{rank}(M)\leq4$ established above. Two heads cannot represent $f_{10}$, proving $3\leq\hstar(f_{10})$ and completing Theorem~\ref{thm:separation}.

\subsection{Weighted-Sum Interpolation}
\label{app:proof-weighted}

We prove Theorem~\ref{thm:weighted-upper}. Set $M=|\operatorname{Im}(t_\lambda)|$. The claim is immediate when $M=1$, so assume $M\geq2$. Let $\Lambda=\sum_i\lambda_i$ and enumerate the image of $t_\lambda$ as $0=\tau_0<\tau_1<\cdots<\tau_{M-1}$.

Choose distinct $\alpha_1,\ldots,\alpha_{M-1}>1$ and define
\begin{equation}
\phi_j(\mathbf{x})=\frac{\alpha_j t_\lambda(\mathbf{x})}{1+\Lambda+(\alpha_j-1)t_\lambda(\mathbf{x})}
=g_j(t_\lambda(\mathbf{x})).
\label{eq:weighted-response}
\end{equation}
The denominator is strictly positive on the cube and has strictly positive slopes $(\alpha_j-1)\lambda_i$. Thus these affine ratios are atoms by the coefficient characterization in Appendix~\ref{app:proof-normal-form}.

It remains to show that $1,g_1,\ldots,g_{M-1}$ form a basis for functions on $\operatorname{Im}(t_\lambda)$. Suppose a linear combination of these functions vanishes on the image. Evaluation at $\tau_0=0$ eliminates its constant coefficient. Dividing the remaining equations by each $\tau_m>0$ and setting $r_j=(1+\Lambda)/(\alpha_j-1)$ reduces them to $\sum_jd_j/(\tau_m+r_j)=0$ for $m=1,\ldots,M-1$. After multiplication by $\prod_j(s+r_j)$, the corresponding rational function has a numerator of degree at most $M-2$ with $M-1$ distinct roots. It is therefore identically zero. Evaluating at each pole $-r_j$ shows that every $d_j$, and hence every original coefficient, vanishes. The $M$ functions are linearly independent on an $M$-point set, so they form a basis.

Interpolating the target signs gives a score $c+\sum_{j=1}^{M-1}a_jg_j(t_\lambda(\mathbf{x}))$. Each scaled term is an atom: multiplying its numerator by $a_j$ preserves admissibility because its denominator has strictly positive slopes. The normal form in Theorem~\ref{thm:normal-form} therefore gives an $(M-1)$-head representation. Taking $\lambda_i=2^{i-1}$ makes $t_\lambda$ injective on the Boolean cube, so $M=2^n$ and $\hstar(f)\leq2^n-1$ for every Boolean function.

\subsection{Exact Symmetric Characterization}
\label{app:proof-symmetric}

We prove Theorem~\ref{thm:symmetric}. Write $\sigma_k=+1$ when $F(k)=1$ and $\sigma_k=-1$ otherwise, and let
\begin{equation}
T=\{k\in[n]:\sigma_{k-1}\neq\sigma_k\},
\qquad |T|=\signchanges.
\end{equation}
The polynomial
\begin{equation}
P(z)=\sigma_0\prod_{k\in T}\left(k-\frac12-z\right)
\label{eq:symmetric-sign-polynomial}
\end{equation}
has sign $\sigma_k$ at every integer $k\in\{0,\ldots,n\}$ and degree $\signchanges$. Thus $P(|\mathbf{x}|)$ sign-represents $f$.

For the reverse threshold-degree inequality, take any degree-$d$ polynomial sign-representing $f$, multilinearize it on the Boolean cube, and average it over all coordinate permutations. The averaged polynomial remains a strict sign representation because $f$ is symmetric. Every symmetric multilinear polynomial of degree at most $d$, evaluated on a Boolean input of weight $k$, is a univariate polynomial $R(k)$ of degree at most $d$: the degree-$m$ elementary symmetric polynomial evaluates to $\binom{k}{m}$. Whenever $F$ changes between $k-1$ and $k$, the values $R(k-1)$ and $R(k)$ have opposite signs, so $R$ has a root in $(k-1,k)$. These intervals are disjoint, giving $d\geq\signchanges$. Hence
\begin{equation}
\tdeg(f)=\signchanges,
\end{equation}
and Theorem~\ref{thm:threshold} gives $\hstar(f)\geq\signchanges$.

For the upper bound, assume $C=\signchanges>0$ and choose distinct $r_1,\ldots,r_C>0$. Set $B(z)=\prod_{h=1}^{C}(z+r_h)$. Since $B(k)>0$ at every Hamming weight, $P/B$ has the target sign pattern. Partial fractions give
\begin{equation}
\frac{P(z)}{B(z)}=c+\sum_{h=1}^{C}\frac{d_h}{z+r_h}.
\label{eq:symmetric-partial-fractions}
\end{equation}
Each shifted reciprocal $d_h/(|\mathbf{x}|+r_h)$ is an atom by Appendix~\ref{app:proof-normal-form}: its denominator is strictly positive on the cube and has all slopes equal to one, while its numerator is constant. Equation~\eqref{eq:symmetric-partial-fractions} is therefore realized with $C$ heads. If $C=0$, $f$ is constant and needs no heads. This proves $\hstar(f)\leq\signchanges$ and completes the equality.

\section{Initial Prompt and Repository README of the Autonomous Run}
\label{app:prompt}

\subsection{Initial prompt}

The single user message that started the theorem-generation run, reproduced from the retained rollout (25 August 2026, 03:48:34 UTC). Typos are as in the original. The counterexample search was opened on 13 July 2026 with the one-line goal \emph{find an explicit constructive example when $\tdeg(f)$ not equal to $\hstar(f)$}; the retained rollout covers that single turn.

\begin{sourcebox}
/goal Answer the main question posed in the \texttt{README.md} file. Here are some additional guidelines that should help you get to the best possible answer:
\begin{itemize}[leftmargin=*,nosep]
  \item Assume for purposes of this task that a complete affirmative result exists. Partial progress does not count unless it implies exactly the resolution above. In particular, proofs for very narrowly restricted boolean function classes, very weak upper or lower bounds, and trivial or simple enough cases are insufficient.
  \item Begin with a genuinely diverse portfolio of approaches. Explore substantially different formulations, invariants, reductions, algebraic or geometric viewpoints, extremal arguments, and computational sanity checks.
  \item Maintain an explicit registry of approach families. Group your experments by the mathematical idea they are using, not by superficial wording. If many attempts converge to one family, redirect some of them toward underexplored formulations.
  \item Do not allow one approach to dominate merely because it gives elegant reductions. A route that ends at a lemma equivalent in strength to the original conjecture is not close to completion unless it supplies a genuinely new proof of that lemma.
  \item When an approach stalls at a theorem-strength missing lemma, mark that route as blocked. Only continue working on it if someone proposes a materially new mechanism, invariant, or construction.
  \item Keep several incompatible proof routes alive through multiple rounds. Cross-pollinate ideas only after you have developed them far enough to expose their real strengths and gaps.
  \item Require results to container concrete theorems, constructions, equations, or counterexamples to previously proposed such. Reject status reports, vague optimism, and claims that an unproved global compatibility statement is ``routine.''
  \item You should repeatedly synthesize, challenge, redirect, and launch new rounds. Do not stop after the first wave fails. Produce a complete proof if one survives audit; otherwise report only the strongest rigorously proved derivation and its exact remaining gap. Do not return merely because current approaches fail or you report theorem-strength gaps. Continue launching new rounds, reopening blocked approaches only when there is a genuinely new mechanism, and searching for fresh formulations. Return only when a complete affirmative proof has been found and survives adversarial audit. Do not return a reduction, partial result, isolated missing lemma, ``best effort'' summary, or explanation of why the problem is difficult. Spend at least 8 hours on this before even thinking of returning or giving up. Public search may not be used for ordinary mathematical background or standard named theorems, neither to search for a solution to this exact conjecture or benchmark.
\end{itemize}
\end{sourcebox}

\subsection{Repository README at the start of the run}
\label{app:readme}

The prompt refers to the main question posed in the repository's \texttt{README.md}. At the start of the run the repository contained only this file, \texttt{problem\_statement.md}, and \texttt{model.md}; the agent rewrote the README during the run, so the version below is the one it was given, reproduced verbatim.

\begin{sourcebox}

  A from-scratch attack on the head-complexity problem: what is the minimum number of attention heads $H^{\ast}(f)$ that a single-layer, attention-only transformer with a linear readout needs to compute a Boolean function $f$?

This repo deliberately starts with only the problem definition. No prior results, proofs, or surveys are carried over.
\begin{itemize}[leftmargin=*,nosep]
  \item \texttt{problem\_statement.md}: the problem, the central quantity $H^{\ast}(f)$, and the core questions.
  \item \texttt{model.md}: the precise architecture (embeddings, attention heads, residual stream, readout, masking convention, deliberate simplifications).
\end{itemize}
\end{sourcebox}

\subsection{Problem statement at the start of the run}
\label{app:problem-statement}

Reproduced verbatim from \texttt{problem\_statement.md}, which did not change during the run.

\begin{sourcebox}
\paragraph{Goal}

We want to characterize the minimum number of attention heads required to compute a Boolean function

\[f : \lbrace 0,1 \rbrace^n \to \lbrace 0,1 \rbrace\]

in a \textbf{single-layer attention-only transformer} with a \textbf{linear readout} from a designated query token.

The motivating example is that, in the 2-bit setting, a single head can linearly realize functions such as $\mathrm{AND}$ and $\mathrm{OR}$, but cannot realize $\mathrm{XOR}$, while two heads suffice. This suggests that the number of heads may define a meaningful complexity measure for Boolean functions.

\paragraph{Model}

We fix the following architecture.

\begin{itemize}[leftmargin=*,nosep]
  \item \textbf{Input:} a sequence consisting of $n$ input bits and one dedicated query token \texttt{=}.
  \item \textbf{Embedding:} token embeddings plus positional embeddings.
  \item \textbf{Computation:} a \emph{single self-attention layer} with $H$ parallel heads and no MLP.
  \item \textbf{Output:} the residual stream at the query token after the attention update.
  \item \textbf{Readout:} a linear probe (equivalently, a thresholded affine functional) applied to that query residual.
\end{itemize}

A Boolean function $f$ is said to be \textbf{computable with $H$ heads} if there exists a choice of embeddings, attention parameters, and linear readout such that the resulting classifier equals $f$ on all inputs in $\lbrace 0,1 \rbrace^n$.

\paragraph{Main Quantity}

Define

\[H^{\ast}(f) := \min \left\lbrace H : f \text{ is computable with } H \text{ heads in the above model} \right\rbrace.\]

Our central problem is to understand $H^{\ast}(f)$ as a function of $f$.

\paragraph{Core Questions}

\begin{enumerate}[leftmargin=*,nosep]
  \item \textbf{Exact characterization.} Can $H^{\ast}(f)$ be expressed, exactly or approximately, in terms of a known invariant of $f$?
  \item \textbf{Lower bounds.} What techniques prove that a function requires at least $H$ heads?
  \item \textbf{Upper bounds.} Can we constructively realize broad classes of Boolean functions with few heads?
  \item \textbf{Comparison to classical complexity measures.} How does $H^{\ast}(f)$ relate to:
\begin{itemize}[leftmargin=*,nosep]
  \item threshold degree,
  \item rational degree,
  \item threshold density / sparsity,
  \item circuit depth / threshold-circuit complexity,
  \item Fourier or spectral complexity measures?
\end{itemize}
  \item \textbf{Natural families.} What is $H^{\ast}(f)$ for standard functions such as:
\begin{itemize}[leftmargin=*,nosep]
  \item $\mathrm{AND}$, $\mathrm{OR}$,
  \item parity / $\mathrm{XOR}$,
  \item majority,
  \item exact-count / symmetric functions,
  \item address / indexing functions,
  \item intersections of threshold functions?
\end{itemize}
\end{enumerate}

\paragraph{Desired Outcome}

We would like to identify either:

\begin{itemize}[leftmargin=*,nosep]
  \item an invariant $I(f)$ such that $H^{\ast}(f)$ is controlled by $I(f)$ (ideally $H^{\ast}(f) \asymp I(f)$ on broad classes), or
  \item explicit function families for which the head complexity exhibits qualitatively new behavior compared to classical measures.
\end{itemize}

In short:

\begin{quote}\textbf{What is the right notion of Boolean-function complexity captured by the minimum number of attention heads in one-layer attention with a linear readout?}\end{quote}
\end{sourcebox}

\subsection{Model specification at the start of the run}
\label{app:model-spec}

Reproduced verbatim from \texttt{model.md}, which did not change during the run.

\begin{sourcebox}
\paragraph{Purpose}

This file fixes the mathematical model used throughout the project.

We study Boolean functions

\[f : \lbrace 0,1\rbrace^n \to \lbrace 0,1\rbrace\]

computed by a one-layer attention-only architecture with a linear readout from a designated query token.

\paragraph{Input Sequence}

For an input

\[x = (x_1, \ldots, x_n) \in \lbrace 0,1\rbrace^n,\]

the model sees the sequence

\[(x_1, \ldots, x_n, =).\]

The final symbol \texttt{=} is a distinguished query token. We only read out the residual stream at that token.

\paragraph{Embeddings}

Fix a model dimension $d_{\mathrm{model}} \geq 1$ and a head dimension $d_{\mathrm{head}} \geq 1$.

Choose token embeddings

\[e_0, e_1, e_= \in \mathbb{R}^{d_{\mathrm{model}}}\]

and positional embeddings

\[p_1, \ldots, p_n, p_= \in \mathbb{R}^{d_{\mathrm{model}}}.\]

For an input $x$, define the embedded vectors

\[u_i(x) := e_{x_i} + p_i \qquad \text{for } 1 \leq i \leq n,\]

and

\[u_=(x) := e_= + p_=.\]

Note that $u_=(x)$ is constant as a function of $x$.

\paragraph{One Attention Head}

For each head $h \in \lbrace 1, \ldots, H\rbrace$, choose matrices

\[W_Q^{(h)}, W_K^{(h)}, W_V^{(h)} \in \mathbb{R}^{d_{\mathrm{head}} \times d_{\mathrm{model}}}.\]

Also choose output-projection blocks

\[W_O^{(h)} \in \mathbb{R}^{d_{\mathrm{model}} \times d_{\mathrm{head}}}.\]

Equivalently, one may concatenate these blocks into a single matrix

\[W_O = \begin{bmatrix} W_O^{(1)} & \cdots & W_O^{(H)} \end{bmatrix} \in \mathbb{R}^{d_{\mathrm{model}} \times H d_{\mathrm{head}}}.\]

For each position

\[j \in \lbrace 1, \ldots, n, =\rbrace,\]

define

\[q^{(h)}(x) := W_Q^{(h)} u_=(x) \in \mathbb{R}^{d_{\mathrm{head}}},\]

\[k_j^{(h)}(x) := W_K^{(h)} u_j(x) \in \mathbb{R}^{d_{\mathrm{head}}},\]

and

\[v_j^{(h)}(x) := W_V^{(h)} u_j(x) \in \mathbb{R}^{d_{\mathrm{head}}}.\]

The query-token logit is

\[\ell_j^{(h)}(x) := \bigl(q^{(h)}(x)\bigr)^\top k_j^{(h)}(x).\]

The attention weights are the softmax probabilities

\[\alpha_j^{(h)}(x) := \frac{\exp\bigl(\ell_j^{(h)}(x)\bigr)}{\sum_{k \in \lbrace 1,\ldots,n,=\rbrace} \exp\bigl(\ell_k^{(h)}(x)\bigr)}.\]

The unprojected output of head $h$ at the query position is

\[\widetilde y^{(h)}(x) := \sum_{j \in \lbrace 1,\ldots,n,=\rbrace} \alpha_j^{(h)}(x) v_j^{(h)}(x) \in \mathbb{R}^{d_{\mathrm{head}}}.\]

The projected contribution of head $h$ to the residual stream is

\[y^{(h)}(x) := W_O^{(h)} \widetilde y^{(h)}(x) \in \mathbb{R}^{d_{\mathrm{model}}}.\]

\paragraph{Residual Stream At The Query Token}

With $H$ parallel heads, the residual stream at the query token after the attention layer is

\[r(x) := u_=(x) + \sum_{h=1}^{H} y^{(h)}(x).\]

This is the only representation used by the final classifier.

\paragraph{Readout}

Choose a readout vector

\[w \in \mathbb{R}^{d_{\mathrm{model}}}\]

and a threshold

\[\tau \in \mathbb{R}.\]

The classifier outputs

\[f(x) = 1 \qquad \Longleftrightarrow \qquad w^\top r(x) > \tau.\]

Equivalently, one may write an affine score

\[S(x) := w^\top r(x) - \tau\]

and classify by the rule

\[f(x) = 1 \qquad \Longleftrightarrow \qquad S(x) > 0.\]

\paragraph{Computability And Head Complexity}

A Boolean function $f$ is \textbf{computable with $H$ heads} if there exist:

\begin{itemize}[leftmargin=*,nosep]
  \item a model dimension $d_{\mathrm{model}}$,
  \item a head dimension $d_{\mathrm{head}}$,
  \item token and positional embeddings,
  \item attention parameters $\lbrace W_Q^{(h)}, W_K^{(h)}, W_V^{(h)}, W_O^{(h)}\rbrace_{h=1}^{H}$,
  \item a readout vector $w$ and threshold $\tau$,
\end{itemize}

such that the resulting classifier agrees with $f$ on every input in $\lbrace 0,1\rbrace^n$.

We then define

\[H^{\ast}(f) := \min \left\lbrace H : f \text{ is computable with } H \text{ heads in the above model} \right\rbrace.\]

\paragraph{Masking Convention}

The formulas above sum over all input positions and the query position itself.

For the current project, this is equivalent to the usual causal-mask convention, because:

\begin{enumerate}[leftmargin=*,nosep]
  \item the query token \texttt{=} is placed last in the sequence,
  \item only the residual stream at that final token is ever read out.
\end{enumerate}

So a standard causal mask would still allow the query token to attend to every input bit and to itself. None of the current results depend on choosing between these two conventions.

\paragraph{Deliberate Simplifications}

The model in this project deliberately excludes several features of a full transformer block:

\begin{itemize}[leftmargin=*,nosep]
  \item no MLP,
  \item no layer normalization,
  \item no dropout,
  \item no multi-layer stacking,
  \item no extra nonlinearity after the attention update.
\end{itemize}

The only nonlinearity in the model is the softmax inside each attention head, followed by the final thresholding in the readout.

\paragraph{Notes On Conventions}

\begin{enumerate}[leftmargin=*,nosep]
  \item Any constant scaling factor in the logits, such as $1 / \sqrt{d_{\mathrm{head}}}$, can be absorbed into the choice of $W_Q^{(h)}$ or $W_K^{(h)}$, so it is omitted from the specification.
  \item Although $W_O$ is now part of the formal model, many lower-bound arguments can compose the final linear readout with $W_O^{(h)}$ and thereby reduce each head again to a scalar contribution.
  \item Because $u_=(x)$ is constant in $x$, any contribution from the query token's skip connection can always be absorbed into the readout threshold when proving lower or upper bounds.
\end{enumerate}
\end{sourcebox}

\end{document}